# CORRELATING FEATURES AND CODE BY DYNAMIC AND SEMANTIC ANALYSIS

Ren Wu

Shanghai Lixin University of Commerce, Shanghai 201620, China

## ABSTRACT

*One major problem in maintaining a software system is to understand how many functional features in the system and how these features are implemented. In this paper a novel approach for locating features in code by semantic and dynamic analysis is proposed. The method process consists of three steps: The first uses the execution traces as text corpus and the method calls involved in the traces as terms of document. The second ranks the method calls in order to filter out omnipresent methods by setting a threshold. And the third step treats feature-traces as first class entities and extracts identifiers from the rest method source code and a trace-by-identifier matrix is generated. Then a semantic analysis model-LDA is applied on the matrix to extract topics, which act as functional features. Through building several corresponding matrices, the relations between features and code can be obtained for comprehending the system functional intents. A case study is presented and the execution results of this approach can be used to guide future research.*

## 1. INTRODUCTION

Software maintenance often involves time-consuming and tedious activities. Maintainers spend 50% up to almost 90% of their time trying to understand the program so as to make changes correctly [1]. To understand the underlying intents of an unfamiliar system, maintainers look for clues in the code and documentation. These clues include program semantic information e.g. identifiers and comments, program structural information such as programming patterns and idioms, program dependencies and so on.

Feature location is to determine the relationships between domain concepts and other software artefact [2], e.g., set of execution methods and the related method declarations in source code. We refer to a feature is a unit of user-observable behaviour of a system. There are various techniques for feature location such as static analysis, dynamic analysis, formal concept analysis etc. In this paper we introduce an approach to analyse dynamic information combining with semantic information. The main reasons for this combination are as follows:

Dynamic analysis: In reverse engineering approaches, it is common to focus on analysing static source code entities of a system. But it is difficult for us to determine what roles for software entities playing in a software system and how these functions interact [3]. A set of trace events generated by exercising desired use cases can be analysed to correlate features and software entities e.g. methods and classes. As execution traces show system's runtime behaviour, so it is helpful for us to understand how the system's functions are implemented by dynamic analysis.

Semantic analysis: Only analysing the structure of a software system to understand program is not enough, since the structure tells us only how the system is working but not what the system is about [4]. The names of identifiers and comments of source code can help understand what the system is about by semantic analysis. A programmer's code and comment often hint the content of design documents and requirement documents, so it is possible to extract semantic information associated with high-level business functions from the comments and identifier names of the source code using information retrieval technology.

We focus on the correlation between features and code using dynamic and textual analysis technology. The assumption is that the maintainer has limited prior familiarity with the target system and wants to know how many features in the system and where they are implemented. A simple solution is to exercise use cases as many as possible. By doing so, on one hand, to help the maintainer know some functions of the target system, on the other hand, to locate features by analysing generated trace events. Rather than assuming a one-to-one correspondence between features and scenarios, we consider a scenario might tangle many features, and a feature scatter in many execution traces. In other words, our approach is not restricted only to start from a single feature and find its corresponding code, but on the assumption that maintainer start from a set of use case scenarios, and then get relevant features, that is, the method is able to extract more features and correlate features and code with a probability distribution.

The paper presents an implementation approach for finding features and locating features in code. To this end, we first start by generating dynamic traces executing usage scenarios. The execution result is the methods execution calls within trace then becomes the document corpus under analysis. Meanwhile, method identifiers are parsed and extracted to establish the corpus of terms and documents for LDA analysis, and the utility functions (i.e. methods with very high frequency counts) are removed from the document corpus. And then LDA is used on the set of topics-to-terms associations. Finally, a series of matrix comparisons are generated (e.g. trace-to-topic, topic-to-identifier, class-by-topic) to calculate semantic similarities. The Luncene tool is used to rank the list of relevant source code fragments, and visualization techniques depicting relevance are also demonstrated using cluster maps.

## 2. BACKGROUND

In this section we will recall necessary basic notions and tools used in our paper for semantic and dynamic analysis proceedings.

### 2.1. Execution Traces

Dynamic information is gathered based on a set of scenarios invoking the related features. A scenario is a sequence of user inputs to a system [5]. As scenarios are being run, trace events are collected. An execution event corresponds to a class or a method to accommodate different levels of granularity.

Figure 1 shows the relations between the execution traces and method entities extracted from the source code. Rather than assuming a one-to-one correspondence between features and scenarios, we assume that a scenario invokes many features. Meanwhile, the features and code units are also many-to-many mappings. We finally obtain a probability distribution between feature and code by LDA extraction, which is more consistent with the actual situation.

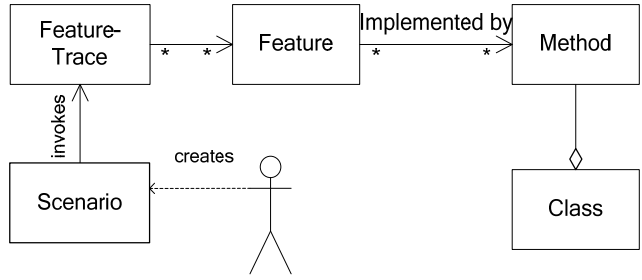

Fig. 1. Feature-Model in UML

In this paper we focus only on method invocation events, but we can also apply to other different levels of abstraction (e.g. classes, packages). We use a profiling tool, namely Java Platform Debugger Architecture (JPDA) [6]. The JPDA-based tracer has the characteristics of flexibility, and allows collect marked-traces during a system's execution so that users can start and stop tracing at will.

We establish the relationships between features and software entities by usage scenarios and capture traces. Based the generation matrix, we cluster and analyse data according to feature criterion. We use a trace-identifier matrix as input for the clustering algorithm. The column represents identifiers extracted from source code and the row is the execution trace. Although a trace might contain many features, however, they are similar on function, so we can apply semantic analysis technology to cluster functional topics on the matrix.

## 2.2. Ranking Execution Traces

A set of traces can be obtained for generating a corresponding matrix: each row represents each trace, and each column represents each method involved in the trace. We use a formula similar to Term-Frequency/Inverse-Document-Frequency to distinguish elements that are specific to a particular trace from those shared by traces. We modify the formula which introduced in [2] to the following scoring formula:

$$socre(m_j) = \sum_i \frac{\text{\# of invoked methods by trace}_i}{\text{\# of total invoked methods}} \times \log\left(\frac{\text{\# traces that invoke method}}{\text{\# traces that invoke method}_{ij}}\right)$$

In the equation, "terms" are methods execution calls within trace, "term frequencies" are number of method calls, and "documents" are executive traces. The algorithm is described in pseudo code in Figure 2.

```
input: method name in each trace; Threshold of score
output: list of score rank
HashMap<String, Float> tf (String[] method) {
    for each method[i] in a trace
      int methodCount = 0;
      for each method[j] in a trace
        if (i != j) {
        if (method[i].equals(method[j])) {
            method[j] = " ";
            methodCount++;
        }
      }
    if (method[i] != " ")
    {
      tf.put(method[i](new Float(++methodCount))/methodNum);
      method[i] = " ";
```

```
        }
    }
        return tf;
}
HashMap<String, Float> idf () {
……
Calculate the number of traces D
Calculate the number of traces that active the current Method Dc
        idf.put(method, Log.log(D / Dc, 2)); }
        return idf;
    }
……
Score(mj) = Sum(tf * idf) of each method
Mt={mj | score (mj) >= Threshold}
```

Fig. 2. The algorithm of traces score

### 2.3. Extracting semantic information

The process steps for extracting semantic information from the source code have also been previously introduced in [7]. The steps work as follows: Extracting semantic information such as comments and identifiers, from each source code element at a desired level of granularity (e.g., package, class, method). Pre-processing the textual information, e.g., perform stemming and remove stop words. Storing the pre-processed data extracted from each code element as a separate document in the document collection.

In this paper we extract the relevant information (e.g., name, parameters, relationships, etc.) from source code by fact extraction and store these data in a relational database. The structure of methods and classes is shown in Table 1.

Table 1. Relevant table structure

| Level of Granularity | Attribute |
|---|---|
| class | ClassName |
|  | InheritsFrom |
|  | ImplemetsTo |
|  | Variables |
| method | MethodName |
|  | Arguments |
|  | ReturnType |
|  | ReturnValue |
|  | Terms in comments |

We collect a set of traces in a corpus, For example, each document in a corpus. The first line is the number of traces, and each successive line is a single trace in the collection with a set of terms or identifiers. In the paper, the list of identifiers from the above database corresponding to a single method name or class name. Because we have stored these identifiers in a database, so we just need to sort related method name and fetch the corresponding data, then write the information to the document. This format is consistent with the input format required by the LDA tool used in textual analysis.

### 2.4. LDA Model

Latent Dirichlet Allocation (LDA) is a document model which explicitly models topic multiplicity of each document. It is a generative probabilistic model of a corpus. The basic idea is that documents are represented as random mixtures over latent topics, where each topic is characterized by a distribution over words [8]. In LDA, documents are represented as mixtures over latent topics, and each topic is characterized by a distribution over words.

The LDA-based model assumes a prior Dirichlet distribution on θ, thus allowing the estimation of φ without requiring the estimation of θ. LDA assumes the following generative process for each document w in a corpus D:

1. Choose N ~ Poisson(ξ).

2. Choose q ~ Dir(α).

3. For each of the N words wi:

(a) Select a topic tk ~ Multinomial (θ).

(b) Select a word wi from p(wi|zw,β), a multinomial probability conditioned on topic tk.

There are more details pertaining to LDA in [8].

### 2.5. Semantic Clustering

To get a semantic model of the software system, we implement these steps as follows:

First, we split the software system into text documents and use the textual representation of traces as a document. Second, we sort methods on the database for obtaining the related attribute values, then we replace terms with identifiers by pair-wise matrix multiplication. An identifier is any word found in the source code or comments, except keywords of the programming language. Identifiers are separated based on standard naming conventions. Considering traces as a document and an identifier as a term，we create a trace-by identifier matrix, similar to the common document-term matrix.

### 2.6. Applying LDA

Since LDA is basically a topic modelling technique, it not only discovers similarity between identifiers, it also creates a cluster of similar identifiers to form a topic. In this paper we use LDA to cluster identifiers according to feature criterion.

Generating probability distributions: Topic-identifier and trace-topic matrices are obtained after applying LDA, in the matrices each trace is probabilistically associated with a set of topics and each topic is probabilistically associated with a set of identifiers.

Grouping similar topics: The process is to calculate cosine similarity among topics to group similar topics for further analysis. Two topics will be clustered into the same category if a cosine similarity between them is greater than a certain threshold [9]. A topic can belong to several different categories, and the result is a list of categories and some rest topics.

## 3. OUR APPROACH

We outline how we apply our technique to incorporate feature-traces model of source code and how to extract topics from the trace model. The approach is a combination coming from two different sources: the execution traces of scenarios, and the comments and identifiers extracted from the class calls or method calls source code. The whole execution process is illustrated in Figure 3.

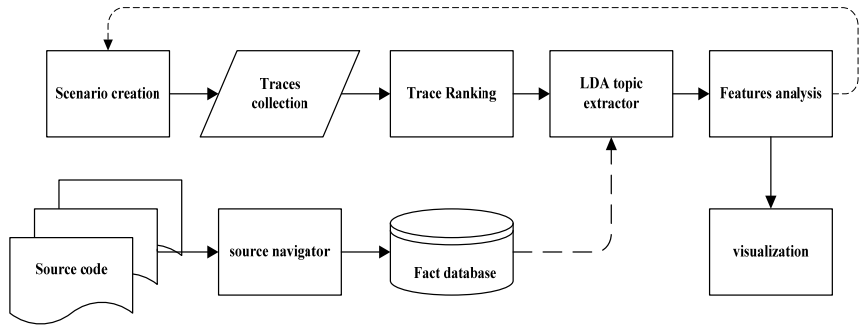

Fig. 3. Overview of our approach

### 3.1. Traces as text corpus

We assume one has limited prior familiarity with the target system, so he or she exercise some scenarios for understanding system's functions by clicking the button or menu on GUI to obtain traces events. JPDA profiling tool is used to extract execution traces and model as a list of methods within traces. The tool can collect marked traces to reduce the size of the traces or collect complete traces without mark at will [10], and the outputs of the tool are a set of events in each thread.

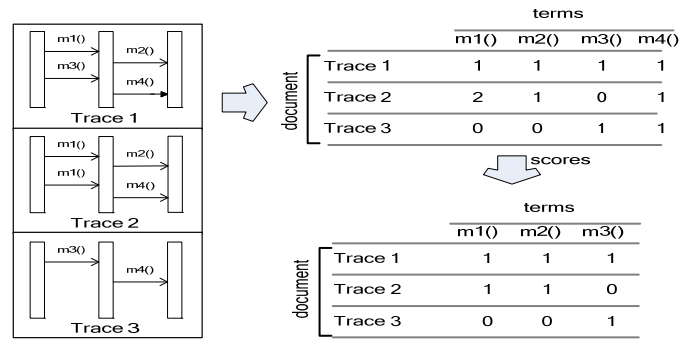

Fig. 4. The generation of a document and terms

We rank the methods of the collection by calculating rank score according to the algorithm in Section 2, and set a threshold to filter out some methods with very high frequency counts, which are useless to special-function. We use a similar example with [11] to show how Traces form a document and the methods form the terms.

Table 2. The score of each method

| tf-idf | m1 | m2 | m3 | m4 |
|---|---|---|---|---|
| Trace1 | 0.044022813 | 0.044022813 | 0.044022813 | 0 |
| Trace2 | 0.08804563 | 0.044022813 | 0 | 0 |
| Trace3 | 0 | 0 | 0.08804563 | 0 |
| Score(m) | 0.132068443 | 0.088045626 | 0.132068443 | 0 |

The score of each method is calculated and show the result is shown in Table 2. Some methods with lower score will be removed. For instance, m4 () is considered as a utility method and is filtered out.

In analysis we consider only the relationships between traces and software entities e.g., classes or methods. For mapping each cell to concept, we keep one occurrence of any repetition only if a concept is repeated many times in a trace. Because a repetition does not introduce a new concept, thus we compress traces to remove repetitions. Therefore, in Figure 4 we change the cell value of m1 () in trace 2 from 2 to 1.

### 3.2. Building the Trace-Identifier Matrix

We treat traces as text corpus and analyse the entities and related static structural model of source code. In this way, the relationships between the method calls in traces and the static model method entities are established and a matrix is generated with each column represents an identifier of source code and each row is a trace. The value of each cell in the matrix is either 0 or 1, indicating the absence or presence of a method or a class in the trace. The generation process of the trace-identifier matrix is shown in Figure 5.

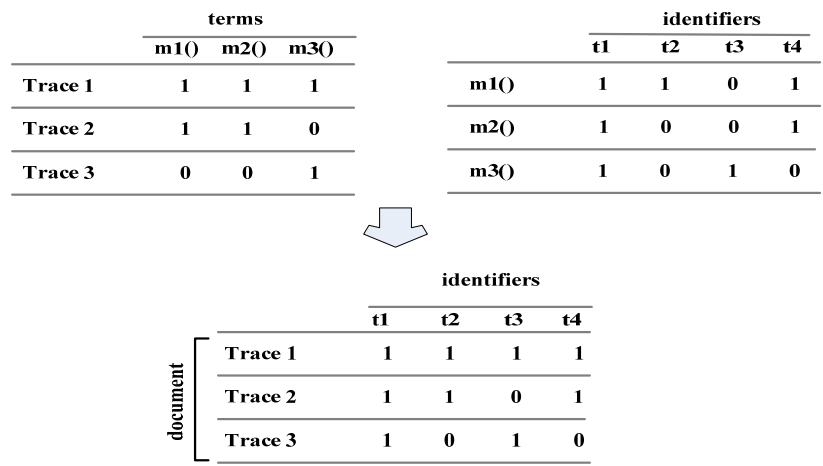

Fig. 5. The generation of trace-identifier matrix

Our semantic analysis tool is applied on the traces. To use the traces as text corpus, we correlate the method names found in the traces and the database tables for obtaining related information, and then replace terms with identifiers by pair-wise matrix multiplication. Figure 5 shows the generation process of matrix.

### 3.3. Performing the LDA Analysis

We use an open-source software tool for LDA analysis called GibbsLDA++ [12]. GibbsLDA++ uses Gibbs sampling to estimate topics from the document collection as well as estimate the word-topic and the topic-document probability distributions. The tool also outputs a list of topics with the top n words in the topic, i.e., the n words that have the highest probability of belonging to that topic, where n is a parameter set for each analysis [18].

Besides, after applying LDA to the matrix of trace-identifier, two relevant probability distributions, i.e., trace-topic and topic-identifier matrices are generated. From the results returned by LDA, the most likely terms in each topic, i.e. the topics with the highest probability, can be examined to determine the likely meaning of the topic.

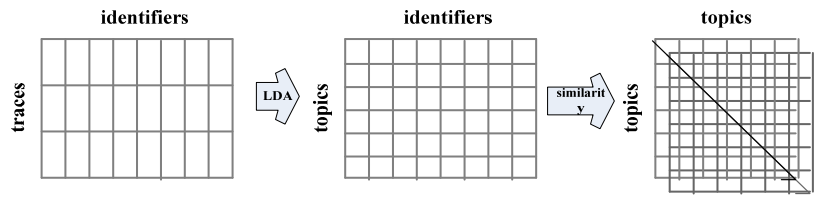

Fig. 6. Clustering by semantic similarity

The trace-topic matrix and the topic-identifier matrix are generated using LDA. Figure 6 shows a partial clustering process. Identifiers can be further clustered into different categories using a certain cosine value based on the corresponding relationships. The semantic similarity shows how two topics are related to each other.

1. CASE STUDY

In this section we present the results of applying our approach to the JHotDraw [13] case study. JHotDraw is an open source tool written in java which provides a graphical user interface. The 6.0 version of the software consists of 486 files representing 650 classes, 4,710 methods, and 845 fields across 28,337 lines of code.

Use cases selection: On the assumption that the requirements information is out-of-date or incomplete, so we exercise execution scenarios as many as possible. The process is to click buttons or menu on the GUI of the system, and our tool collects all trace events to analyse. Table 3 shows a part selected ten use cases [14] in JHotDraw system.

Table 3. Selected use cases in jhotdraw

| U1. Use the Rectangle button to draw a new Rectangle figure |
| U2. Use the Round Rectangle button to draw a new figure |
| U3. Use the ellipse button to draw a new ellipse figure |
| U4. Use the polygon button to draw a new polygon figure |
| U5. Use the line button to draw a new line figure |
| U6. For each graphical element in the load file: turn tracing off; select it; turn tracing on; move it |
| U7. For each graphical element in the load file: turn tracing off; select it; turn tracing on; delete it |
| U8. For each graphical element in the load file: select it; use bring-to-front command from the zoom figure. |
| U9. Use the URL button to attach an URL to each graphical element; to modify the URL on each graphical |
| U10. Create a new text field using the text button; modify the existing text field using the text button |

Table 4 depicts the experimental results of execution traces extraction. The second column shows the number of scenarios in terms of various use cases, the next is the count of execution events in traces. The rest number of methods after filtering is listed in the last column.

Filtering out Omnipresent methods: Software system contains some components that act as mere utilities. Some method invocations related to system events, are not feature-specific and can appear almost anywhere in a trace. Therefore, we decide to remove the omnipresent components in the system. The detection approach has been discussed in Section 2.2. We calculate the trace score and set a certain threshold to filter out utilities. Figure 7 shows the dual axis chat views of the data in Table 4.

Table 4. Results of execution trace extraction

| Use case | Number of scenarios | Number of methods | Number of methods after filtering |
|---|---|---|---|
| U1 | 4 | 7889 | 6657 |
| U2 | 4 | 5040 | 4980 |
| U3 | 5 | 8493 | 7320 |
| U4 | 7 | 10769 | 8960 |
| U5 | 4 | 4253 | 4190 |
| U6 | 4 | 4930 | 4790 |
| U7 | 4 | 5739 | 5432 |
| U8 | 6 | 11078 | 9590 |
| U9 | 5 | 10356 | 10132 |
| U10 | 4 | 6074 | 5879 |

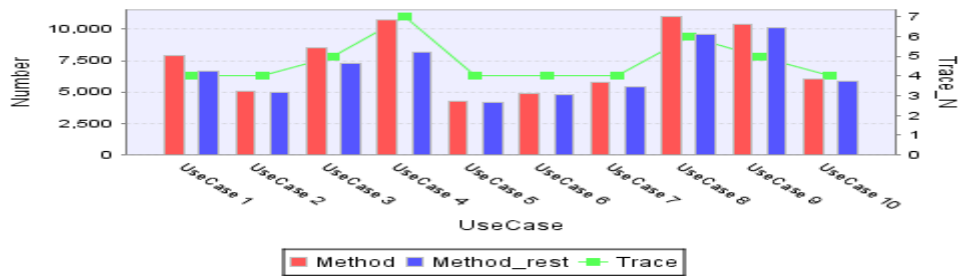

Fig. 7. Dual axis chat views of data in Table III

Applying LDA: The LDA-based approach can capture more subtle statistical relationships among topics, words, entities and traces [15]. We extract 30 topics from the trace-identifier matrix, and list a partial topics and labels in Table 5.

Table 5. A partial extracted topics from jhotdrawtem

| Topic label(feature) | Word list (topic) |
|---|---|
| ITERATION | "list has iterator next add" |
| UNDO | "undo change figure activity affect" |
| DRAW FIGURE | "figure change listener event remove" |
| EVENTS, TOOL | "mouse event tool add down" |
| FIGURE ATTRIBUTES | "attribute figure object constant draw" |
| MAP | "all map contain object vector" |
| DRAW,COLOR | "colour draw print map name" |
| DRAW SHAPE | "point line box draw bound" |
| CONTENT | "content class string producer frame" |
| DRAW,FIGURES | "figure add draw decorated find" |
| DRAW RECTANGLE | "shape rectangle draw rectangular figure" |
| CONNECT | "connection connector figure start end" |
| ROTATION | "radius angle reset rotation view" |
| EVENT, OBSERVER | "desktop component container event listener" |

| RENDERING | "graphic draw image rectangle fill" |
|---|---|
| EDITOR,ANIMAT ION | "editor draw create animation tool" |
| PERSISTENCE | "store read write input output" |
| MOVE | "drag drop source listener target" |

To perform a comparison, we inspect the trace-topic matrix with the probability distribution of topics over traces. For each latent topic identified by a set of 5 words, we assign a meaningful feature name. In some cases the identifiers in a set of words are good enough for identifying a feature.

For example, in Table 5 we see the "draw" features given by "point line box draw bound", "graphic draw image rectangle fill", "draw shape rectangle regular figure", as well as a presence of "iteration" feature with "list iterator has next add" and the "undo" feature given by "undo change figure activity affect" etc.

Querying topics: Because we have generated several mapping matrices such as trace-topic, topic-identifier and class-topic etc., we can find the corresponding relationships among files, components, topics and traces according to the results obtained to each trace. We save these data in database tables, and then use Luncene to index relevant fields in the tables. We have developed a prototype tool with a topic (or feature) querying interface for displaying desired features and their relationships. Figure 8 shows the GUI of our query tool.

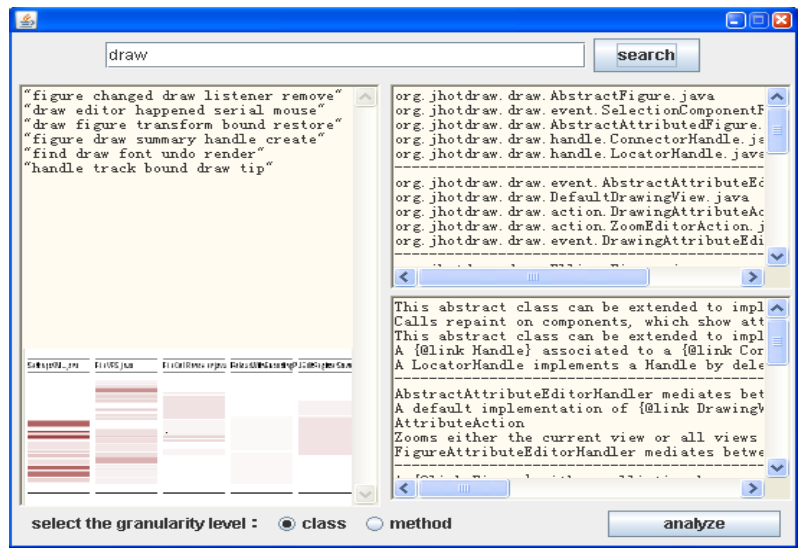

Fig. 8. Our prototype tool for querying

For example, the user formulates a query, which describes the desired features. Our tool sorts the query words from the database. If same words or similar words are present, then the tool outputs all relevant topics in the left panel. For each topic in the list, the user can click it, and then the corresponding classes or methods names are listed in the right panel. The user can further investigate the relevant feature-traces events, and a graphical view of the topic probability distribution of the files is also shown in the GUI.

Since features are not the primary units of decomposition in object-oriented programs, so it is difficult for implementations of features to be represented explicitly in programs' source code. However, we are able to identify this relationship by using probability distribution, which is more consistent with the actual situation. Table 6 lists the set of components and associated probabilities that related to a topic.

Table 6. Corresponding probability values of components for sample topics

| topic | class | Probability |
|---|---|---|
| "draw, editor, happen, serial, mouse" | AbstractAttributeEditorHandler.java | 0.177966 |
| | DefaultDrawingView.java | 0.127586 |
| | DrawingAttributeAction.java | 0.122642 |
| | ZoomEditorAction.java | 0.119565 |
| | DrawingAttributeEditorHandler.java | 0.101153 |
| "draw, figure, transform, bound, restore" | EllipseFigure.java | 0.383333 |
| | RectangleFigure.java | 0.382609 |
| | ImageFigure.java | 0.378571 |
| | AbstractAttributedDecoratedFigure.java | 0.373134 |
| | TextAreaFigure.java | 0.368421 |
| **topic** | **method** | **Probability** |
| "draw, stroke, fill, view, decorate" | DrawingEditor.createInputMap() | 0.270270 |
| | BezierControlPointHandle.draw(Graphics2D) | 0.153846 |
| | BoundsOutlineHandle.draw(Graphics2D) | 0.152941 |
| | drawFigure(Graphics2D) | 0.152778 |
| | ConvexHullOutlineHandle.draw(Graphics2D) | 0.146067 |
| "draw, composite, scale, image, hint" | ODGGroupFigure.draw(Graphics2D) | 0.379310 |
| | SVGPathFigure.draw(Graphics2D) | 0.337079 |
| | ODGPathFigure.draw(Graphics2D) | 0.314607 |
| | SVGAttributedFigure.draw(Graphics2D) | 0.314376 |
| | SVGGroupFigure.draw(Graphics2D) | 0.310345 |

Generating categories: To group similar topics on topic-identifier matrix, we compute cosine similarity between each pair of topics. If a cosine similarity between two topics is greater than a certain threshold, we cluster them into the same category [9]. Some similar features are clustered together in terms of similarity comparison. For example, we set the threshold is 0.6, the similarity of "tool enabled object usable palette" and "editor drawing create animation tool" is above 0.6, which means the draw tool for use; "drag drop target source listener" and "event listener object fire iterator" show listener event happening; "invoke handle owner step undo" and "undo figure figures activity affected" show undo design model etc. By grouping relevant files into categories, we further comprehend the system functional intents.

Forming class-topic matrix: We can also examine the relationships between topics and files by the class-topic matrix. Table 7 shows a partial probability distribution data selected from the JHotDraw experiment results.

Table 7. A partial of probability distribution in JHotDraw

| Class | Topic1 | Topic2 | Topic3 | Topic4 |
|---|---|---|---|---|
| | "mouse event down tool added" | "list next has iterator event" | "mouse event down tool added" | "graphics draw image rectangle fill" |
| AddTool | 0.352631 | 0.243412 | 0.237861 | 0.012321 |
| PolyLineFigure | 0.123021 | 0.534300 | 0.030045 | 0.320000 |
| PolyLineConnector | 0.000120 | 0.650132 | 0.120000 | 0.400521 |
| PolygonTool | 0.000521 | 0.670120 | 0.064521 | 0.414211 |
| PaletteButton | 0.214211 | 0.124832 | 0.214211 | 0.300000 |
| TextAreaFigure | 0.400000 | 0.517232 | 0.490000 | 0.135432 |
| DragTracker | 0.135462 | 0.111213 | 0.567262 | 0.520000 |
| NestedCreationTool | 0.342401 | 0.000000 | 0.220000 | 0.124221 |
| TriangleFigure | 0.526342 | 0.131212 | 0.652123 | 0.004211 |
| FigureChange | 0.621001 | 0.106423 | 0.323232 | 0.000000 |
| FigureChangeAdapter | 0.600000 | 0.006423 | 0.124242 | 0.122462 |

The above data can be graphically displayed using a cluster map view of topic-class, which shows the topic assignment across each component. Figure 9 shows the view, where a class with a higher topic assignment is indicated with a darker colour and vice versa. In this paper, the heat-map technique [16] for graphical representation of data depicting relevance is used, and the clustering visualization is also demonstrated to quickly contrast and compare the distribution in the various topics.

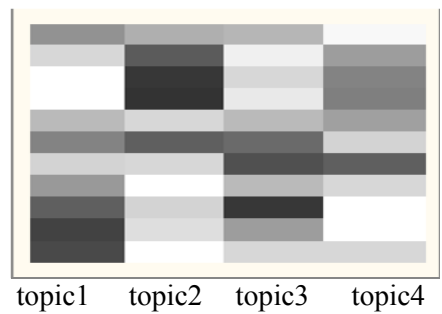

topic1   topic2   topic3   topic4

Fig. 9. Graphical display the relationships of class-topic

We can further cluster these classes by their similarity according to the data in Table 7.

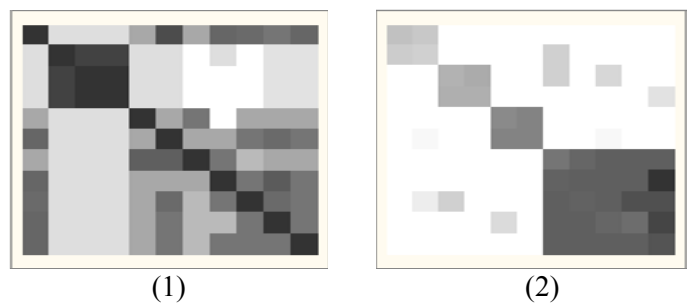

(1)                                  (2)

Fig. 10. Graphical display the similarity of classes

We process the data in Table VII to generate an equivalent matrix. The matrix is graphically displayed in Figure 10 (1). Then we cluster these classes by their similarity by selecting $\lambda$ =0.912 as cut value which determines the following four clusters, and the clusters are displayed in Figure 10 (2).
Cluster 1: class 1, class 6, class 8, class 10, class 11
Cluster 2: class 2, class 3, class 4

Cluster 3: class 9
Cluster 4: class 5, class 7

## 3. DISCUSSION

In this paper we first exercise the system to collect execution traces for some scenarios related to various use cases, which containing sets of different features. Then we compress the traces to remove utilities. Second, we treat traces as first class entities and extract identifiers from the method source code and generate a trace-identifier matrix. Then we apply a semantic analysis tool on the matrix for extracting topics, which as functional features. Our approach makes the syntax and semantics have a good combination.

F-measure: We assume that the maintainer exercises features by clicking the button or menu of the target system for obtaining trace events. Therefore, our feature model does not achieve 100% coverage of the system. To locating more features, we need to increase the coverage the application by exercising more scenarios. However, by extracting identifiers from components source code, expanding the scope of application semantics, to some extent this way can increase the recall. Meanwhile, owing to filter out some concepts useless to system special-functions, we get relative precision for features extracted in the traces. The future work we will compute the F-measure, the harmonic mean of precision and recall: f-measure = 2· (recall · precision) / (recall + precision) by some detail experiment data, and select an acceptable trade-off.

Effectiveness Measure: We build text corpus from the names of the methods involved in the traces, then we sort methods in the database to obtain their corresponding identifiers of source code by our parser tool. We consider it is beneficial for enriching semantic information by add identifiers into analysis instead of methods name only. However, if a great deal identifiers are introduced into analysis models, it is possible to introduce some not feature-specific terms into analysis. So we need to have a choice of variable identifiers.

Topic and Software Feature: Pierre Baldi et al. [15] propose to unify the concept of latent topic with the concept of concern in the domain of software. They claim the distribution of a topic across modules indicates whether the topic is more or less scattered. Their conclusion is a concern is a latent topic. Based on this idea, we focus on features of concerns: our approach obtains concerns from execution traces, filters out some common concepts, and complements semantic concepts into analysis. So we think a feature is also a latent topic. This is a theory of our approach.

## 4. RELATED WORK

Hybrid feature location leverages the benefits of static and dynamic analyses. T.Eisenbarth et al. [5] develop a technique that applies formal concept analysis to traces to produce a mapping of features to methods. A feature-driven approach is presented in [17] by extracting execution traces to achieve an explicit features location. In contrast, our approach not only takes into account mappings between features and software entities like methods or classes, but also extracts more features to help program comprehension.

Marcus et al. [10] present an approach with a single execution trace combining with LSI to extract code relevant to a feature. Adrian Kuhn et al. [11] propose to combine LSI-based analysis and dynamic analysis to identify related features. In contrast, our approach is not restricted only to start from a single feature and find its corresponding code, but on the assumption that maintainer starts from a set of use case scenarios, then get relevant features. The tool for us to use is LDA instead of LSA, because LDA is able to extract features and correlate features and code with a probability distribution, which is more consistent with the actual situation. Besides, a series of matrix comparisons are generated (e.g. trace-to-topic, topic-to-identifier etc.) by LDA tool to calculate semantic similarities.

Several works [9, 18] have shown that the idea of applying LDA in source code for extracting topics can be carried out. In contrast, our approach is not used directly to examine source code for topics, but on the concepts after dynamic processing and filtering, so obtained topics via LDA have functional intents. Besides, Abram et al. [19] treat CVS commits as documents, and compare how extracted topics changed between time periods. LDA has been also used for traceability link recovery [20] and compared to other similar Information Retrieval based approaches, but in [20], topics extraction from requirement documents and design documents not from source code as ours.

## 5. CONCLUSIONS AND FUTURE WORK

Reverse engineering approaches focus only on the implementation details and static structure often ignores the semantics of the problem and solution domain. But semantic information is essential in revealing full of the picture of software system. Our approach discusses a technique for determining which components source code implement which features of software application by incorporating dynamic traces and semantic analysis via a series of pair-wise comparison matrices. We demonstrate the application of the proposed technique via a case study. The main idea behind the technique is to use LDA to extract topics, i.e. functional intents, from methods and identifiers involved in execution traces. The technique is used for feature location in source code.

In the future, we will focus on the following issues: (1) extending the scope of our feature approach to consider exercising more scenarios according to feedback information (2) combining dynamic analysis with system's structural model, e.g. program-element dependencies for enriching the semantic information, (3) assigning automatically meaningful labels to traces to help maintainers understand their meanings, and (4) conducting systematic empirical studies on the choice of the appropriate parameters for the proposed approach and collecting more empirical data to evaluate the effectiveness based on LDA for features location.

**Authors**

**Ren Wu** Received his PhD degree in computer science in 2014. His current research interests include information retrieval, software engineering and software maintenance.

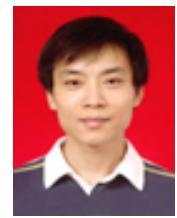